# Triggering of large earthquakes is driven by their twins

Shyam Nandan[1], Guy Ouillon[2] and Didier Sornette[3,4]

[1]*Swiss Seismological Service, ETH Zürich, Sonneggstrasse 5, 8092, Zürich, Switzerland*
*e-mail address: shyam4iiser@gmail.com*
[2]*Lithophyse, 4 rue de l'Ancien Sénat, 06300 Nice, France*
[3] *ETH Zürich, Department of Management, Technology and Economics,*
*Scheuchzerstrasse 7, 8092, Zürich, Switzerland*
[4]*Institute of Risk Analysis, Prediction and Management (Risks-X), Academy for Advanced Interdisciplinary Studies,*
*Southern University of Science and Technology (SUSTech), Shenzhen, China*

Fundamentally related to the UV divergence problem in Physics, conventional wisdom in seismology is that the smallest earthquakes, which are numerous and often go undetected, dominate the triggering of major earthquakes, making prediction of the latter difficult if not inherently impossible. By developing a rigorous validation procedure, we show that, in fact, large earthquakes (above magnitude 6.3 in California) are preferentially triggered by large events. Because of the magnitude correlations intrinsic in the validated model, we further rationalize the existence of earthquake doublets. These findings have far-reaching implications for short-term and medium-term seismic risk assessment, as well as for the development of a deeper theory without UV cut-off that is locally self-similar.

Earthquakes are the sudden bursts of energy release occurring intermittently over an impressive range of scales in the complex spatio-temporal self-organised faulted Earth crust. A common understanding is that earthquakes interact with each other [1] and that earthquake triggering is a crucial part of the organization of seismicity [2]. However, common wisdom in the field holds that small earthquakes, which are numerous and often undetected, dominate the triggering of large events [3-4], requiring the introduction of a UV cut-off to regularize the theory, in the form of a minimum magnitude $M_0$ below which no triggering occurs [5]. Here, we demonstrate that a significant fraction of large events are triggered by previous large events. Being large, these triggering events are reported in available earthquake catalogs. Thus, forecasting large events preceded by a set of seismic precursors would not be spoiled by the noise created by the myriad of small events whose recording is incomplete or whose parameters are highly uncertain. Moreover, our results point to a deeper theory without UV cut-off that is locally self-similar, in the spirit of [6,7].

**Epidemic Type Aftershock Sequence (ETAS) models.** We use the family of ETAS models, which belongs to the class of self-excited Hawkes conditional point processes widely adopted in recent years to model bursty behavior in many systems [8-12], which describe conditional seismicity rate of seismicity as a linear superposition of a stationary background rate, thought to be mainly controlled by the far-field loading dynamics, and a set of transient clusters that are triggered by all past events. In its standard form, the ETAS model describes the conditional seismicity rate of magnitude $m$ events, $\lambda(t, x, y, m | \mathcal{H}_t)$, at any location $(x, y)$ and time $t$ as:

$$\lambda(t,x,y,m|\mathcal{H}_t) = \left[\mu + \sum_{i:t_i<t} g(t-t_i, x-x_i, y-y_i, m_i)\right]\beta e^{-\beta(m-M_0)} \quad (1)$$

where $\mu$ is the background intensity function, assumed to be independent of time, while $\mathcal{H}_t = \{(t_i, x_i, y_i, m_i): t_i < t\}$ stands for the history of the process up to time $t$. The variables $(t_i, x_i, y_i, m_i)$ respectively correspond to the time, x-coordinate, y-coordinate, and magnitude of the $i^{th}$ earthquake in the catalog, while $g(t-t_i, x-x_i, y-y_i, m_i)$ is the triggering kernel, defined in Eq. (2), quantifying the temporal and spatial influence of past events onto future events:

$$g(x,y,t-t_i,x-x_i,y-y_i,m_i) =$$
$$Ke^{a(m_i-M_0)} \frac{T_{norm}\, e^{-(t-t_i)/\tau}}{\{t-t_i+c_0\}^{p_0}} \frac{S_{norm}}{\{(x-x_i)^2+(y-y_i)^2+de^{\gamma(m_i-M_0)}\}^{1+\rho}} \quad (2)$$

$M_0$ is the magnitude of the smallest event able to trigger some other ones, while $T_{norm}$ and $S_{norm}$ ensure normalization of the temporal and spatial components of the triggering kernel, respectively. Eq.(2) combines in a standard way the fertility law $Ke^{a(m_i-M_0)}$ giving the number of events directly triggered by an earthquake of magnitude $m_i$, a time kernel based on the exponentially tapered Omori-Utsu law for aftershocks and a spatial kernel.

The ubiquitous negative exponential Gutenberg-Richter (GR) law, which quantifies the distribution of earthquake magnitudes with exponent $\beta$, is factored out in Eq. (1). This assumes the two following hypotheses: (i) the magnitudes of background and triggered earthquakes are distributed according to the same GR law $e^{-\beta(m-M_0)}$, so that both spontaneous and triggered events differ only in their space and time rates, not in their respective physical origins; (ii) the magnitude distribution of the triggered earthquakes does not depend on the magnitude of the earthquake that triggered them, so that the sole memory of the size of past events is conveyed by the total seismicity rate. Thus, if the space-time rate of immediate future events can be forecasted, their magnitudes cannot be predicted better than by a purely stochastic sampling of the GR law. In most of the literature on the ETAS model, the above two assumptions are tacitly or explicitly accepted. An interesting consequence of the standard ETAS formulation concerns the collective properties of triggering: if $a - \beta < 0$, most earthquakes are triggered by previous small events, and this has been the dominant paradigm until now, based on calibrations of the ETAS model and on other indirect approaches.

Based on recent empirical evidences [13-18], we propose a suite of four models that generalize ETAS and are tested against it (see Supplementary Information (SI), Text S1, for the detailed description of the models):

- Model 1 uses the conditional seismicity rate given by Eq. (1) but extends the standard ETAS model with a space-varying background intensity function [17], with the guiding idea that the future background earthquakes occur mostly in regions where the intensity of past background earthquakes has been high.
- Model 2 is the same as Model 1, but with a modified time kernel $T_{norm}\{t - t_i + c(m_i)\}^{-p(m_i)} e^{-\frac{t-t_i}{\tau}}$, where $c(m_i) = c_0 10^{c_1 m_i}$ and $p = p_0 + p_1 m_i$ [13,18,19]. Thus, the regularizer and the exponent of the time kernel feature exponential and linear dependence on the magnitude of the mainshock, respectively.
- Model 3 is defined by:

$\lambda(t,x,y,m|\mathcal{H}_t) = \mu(x,y)f_{bkg}(m) + \sum_{i:t_i<t} g(t-t_i, x-x_i, y-y_i, m_i)f_{aft}(m|m_i)$ **(3)**

and differs from Model 1 as it features two different size distributions: $f_{bkg}(m)$ is the magnitude distribution of the background earthquakes, following a standard GR form, while $f_{aft}(m|m_i)$ is the conditional distribution of events of magnitude $m$ directly triggered by an earthquake of magnitude $m_i$, according to [6,7,17]. It takes the form of a continuous kinked GR law, with exponent $\beta_1$ for $m < m_i$ and exponent $\beta_2$ for $m \geq m_i$, with $\beta_2 > \beta_1$, so that triggered earthquake magnitudes are more clustered around that of their parent earthquake.

- Model 4 combines all previous novel ingredients of models 2 and 3, i.e., uses different magnitude distributions for triggered events and their trigger as in Model 3, and a time kernel dependent on the magnitude of the trigger as in Model 2.

All four models are calibrated using the EM algorithm proposed by [20], with some necessary modifications allowing for the inversion of the optimal space varying background rate for all models, magnitude dependent parameters of the time kernel (Models 2 and 4) and trigger dependent magnitude distribution (Models 3 and 4). The details of the calibration process are given in the SI, Text S2. The predictive skills of the four models are then evaluated on real data, using earthquakes reported in the Advanced National Seismic System (ANSS) catalog around the state of California, for which we estimated the completeness magnitude $M_c=3$ (see SI, Text S3 for the description of the dataset used in this study, and Text S4 of SI for the testing protocol).

Using rigorous pseudo-prospective cross-validation (i.e., performing out-of-sample predictions of future seismicity in back-tests), we establish that Model 4 is the best model. It outperforms all the other three models in 24 out of 25 test settings (see SI, Text S4). This can be attributed to two crucial properties: a time Omori kernel dependent on the magnitude of the trigger and a kinked GR law conditional on the magnitude of the trigger. The first property has been predicted theoretically and later confirmed by various empirical tests due to the interplay between long-range elastic stress transfer and effective thermal activation of rupture [13,19]. The second property is also a theoretical prediction when imposing a scale-invariant property generalizing the ETAS model so that the minimum cut-off magnitude $M_0$ is pushed to $-\infty$ [6,7]. The first empirical confirmation of the kinked GR law was reported by [16], who proposed a mechanical interpretation using the fact that the symmetry of the deformation tensor at any scale tends to mimic the orthorhombic symmetry of the loading stress field. Thus, each time an event occurs on a fault and creates a monoclinic strain perturbation, another event of similar size tends to be induced on a conjugate fault to re-establish an orthorhombic symmetry.

**What magnitude range dominates triggering?** As Model 4 provides the best available description of seismicity, we are now in position to re-examine the claim that small earthquakes dominate the triggering of large earthquakes [3-4], which is deeply embedded in the structure of the standard ETAS model and its generalization with a spatially heterogenous background rate in Model 1. We address this question by computing the relative contribution $F(m/M)$ to the triggering of events of magnitude larger than $M$ by events of magnitude $m$, for various $M$'s, that Model 4 predicts. For the application to California, we take the average values of parameters inverted using Model 4 on the ANSS catalog: $a = 1.1$, $\beta_1 = 1.6$, $\beta_2 = 3.1$, $\beta_{bkg} = 2.4$, $M_0 = 3$.

Denoting $m$ (resp. $M$) the magnitude of the triggering (resp. triggered) earthquake, for Model 4, the number $N_{>M}^{m>M}(m)$ of events of magnitudes larger than $M$ triggered by an event of magnitude $m$, with $m>M$, is given by

$$N_{>M}^{m>M} = K\,e^{a(m-M_0)}\left[1 - \beta_2 \frac{1}{1-\left(1-\frac{\beta_1}{\beta_2}\right)e^{-\beta_1(m-M_0)}}\left(1-e^{-\beta_1(M-M_0)}\right)\right]$$
$\forall m > M$ **(4)**

which is Eq. (S8-6) in the SI Appendix Text S8, with the correspondence of notations $m_{aft} \to M$ and $m_{main} \to m$.

The number $N_{>M}^{m\leq M}(m)$ of events of magnitudes larger than $M$ triggered by an event of magnitude $m$, with m $\leq$M, is given by

$$N_{>M}^{m\leq M}(m) = K\,\beta_1\,\frac{e^{(a+\beta_2-\beta_1)(m-M_0)}}{1-\left(1-\frac{\beta_1}{\beta_2}\right)e^{-\beta_1(m-M_0)}}\,e^{-\beta_2(M-M_0)} \quad \forall\, m \leq M \quad (5)$$

which is Eq. (S8-10) in the SI Appendix Text S8, with the correspondence of notations $m_{aft} \to M$ and $m_{main} \to m$.

Combining the GR distribution of the background events with these "renormalized" productivity laws for triggered events of magnitudes larger than $M$, the expected contribution of the triggers of magnitudes falling in the interval $[M_0 + i\Delta m, M_0 +$

$(i+1)\Delta m])$ to the triggering of events with magnitude $\geq M$ is given by:

$$E(M_0 + i\Delta m, M_0 + (i+1)\Delta m | M) =$$
$$\begin{cases} \int_{M_0+i\Delta m}^{M_0+(i+1)\Delta m} N_{>M}^{m>M}(m) * \beta_{bkg} e^{-\beta_{bkg}(m-M_0)} dm & \forall\, m > M \\ \int_{M_0+i\Delta m}^{M_0+(i+1)\Delta m} N_{>M}^{m\leq M}(m) * \beta_{bkg} e^{-\beta_{bkg}(m-M_0)} dm & \forall\, m \leq M \end{cases} \quad (6)$$

Then, the fraction $F(m|M)$ of events (of magnitudes $\geq M$) triggered by events of magnitude in the interval $(m, m+\Delta m)$ is obtained by dividing Eq. (6) by the total contribution of earthquakes of all the magnitudes, i.e., by normalizing Eq. (6) by its sum over all triggers' magnitude intervals. Figure 1a shows the resulting $F(m|M)$ for different values of $M$. Except for $M$ values close to the lower boundary $M_0$, one can observe the existence of a secondary peak that becomes dominant for larger values of $M$: the main triggers of large events of magnitude $\geq M$ are events of magnitude $M$. This phenomenon is the more pronounced, the larger $M$ is. This result for $F(m|M)$ is fundamentally different from the prediction of the standard ETAS model, which is given by the blue curve for $M=M_0=3$: in this case, Eq. (4) reduces to $N_{>M}^{m>M}(m) = K\, e^{a(m-M_0)}$, which together with the top expression in Eq. (6) recovers the standard ETAS dependence $\sim K\, e^{(a-\beta_{bkg})(m-M_0)}$. Thus, earthquakes are mostly triggered by earthquakes of similar or larger magnitudes, implying a fundamental change of paradigm in earthquake prediction. The previous paradigm that small earthquakes dominate triggering only holds for triggered earthquakes of magnitudes close to the triggering threshold $M_0$.

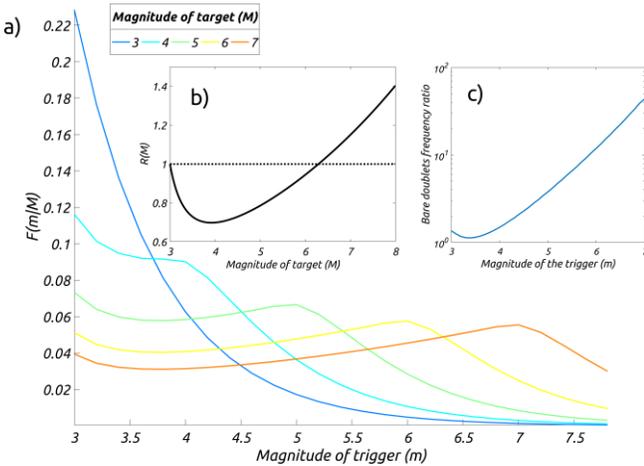

*Fig. 1: **a)** Fraction $F(m|M)$ (obtained from Eq.(6) by normalization) of events (of magnitudes $\geq M$) triggered by events of magnitude in the interval $(m, m+\Delta m)$; Different colors correspond to different values of $M$, indicated in the legend; The curves are obtained for the parameters of Model 4 given by $K=0.45$, $a=1.1$, $\beta_1=1.6$, $\beta_2=3.1$, $\beta=2.4$, $M_0=3$ and $\Delta m=0.2$, which best fit the ANSS Californian catalog. **b)** Dependence of the ratio $R(M)$ given in Eq. (11) as a function of the magnitude of the target, M. **c)** ratio of the number of bare doublets obtained using Model 4 to that same number using Model 1.*

The message of Figure 1a can be complemented by considering the probability $P(m|M)$ that a given event of magnitude $M$ has been triggered by a previous event of magnitude $m$, which reads, using Bayes theorem:

$$P(m|M) = \frac{P(M|m)P(m)}{\int_{M_0}^{+\infty} P(M|m)P(m)dm} \quad (7)$$

where $P(M|m) \sim K e^{a(m-M_0)} f_{aft}(M|m)$ where $f_{aft}(M|m)$ is a kinked GR law of parameters $\beta_1$ and $\beta_2$, defined by Eq. (S8-1) and (S8-2) in Text S8, and $P(m) = f_{bkg}(m)$ is a standard GR law (Eq. S1-3) with parameter $\beta_{bkg}$.

In the case where $m < M$, we have:
$$P(m|M) \sim \beta_1\beta_2 e^{-\beta_2(M-M_0)}\left[e^{(\beta_{bkg}-a-\beta_2)(m-M_0)}\left(\beta_1\beta_2(e^{\beta_1(m-M_0)}-1)\right)\right]^{-1} \quad (8)$$

from which we get, neglecting positive pre-factors:
$$\frac{d}{dm}P(m|M) \sim \beta_2(a+\beta_2-\beta_1-\beta_{bkg})\, e^{\beta_1(m-M_0)} - (\beta_2-\beta_1)(a+\beta_2-\beta_{bkg}) \quad (9)$$

For the parameters calibrated on the Californian catalog with Model 4, this gives $\frac{d}{dm}P(m|M) \sim 0.62\, e^{1.6(m-3)} - 2.7$. Thus, $P(m|M)$ is an increasing function of $m$ for $m \gtrsim 3.9$, and a decreasing function for $m \lesssim 3.9$.

In the case where $m > M$, we have
$$P(m|M) \sim \beta_1\beta_2 e^{-\beta_1(M-M_0)}[\beta_1 e^{(\beta_{bkg}-a-\beta_1)(m-M_0)} + \beta_2 e^{(\beta_{bkg}-a)(m-M_0)}(1-e^{-\beta_1(m-M_0)})]^{-1} \quad (10)$$

After a little algebra and omitting all positive pre-factors, we finally find that $\frac{d}{dm}P(m|M) \sim \beta_2(a-\beta_{bkg})\, e^{\beta_1(m-M_0)} - (\beta_2-\beta_1)(a+\beta_1-\beta_{bkg})$ which takes the form $\frac{d}{dm}P(m|M) \sim -4\, e^{1.6(m-3)} - 0.45$, which is always negative.

Putting these results together, as a function of $m$, $P(m|M)$ is first decreasing for $3 < m < 3.9$, then increasing for $3.9 < m < M$, and then decreasing again for $m > M$. For $3 < M < 3.9$, $P(m|M)$ is monotonically decreasing with m. This implies that, apart from contributions at magnitudes close to $M_0$, the most likely triggering event of an earthquake of magnitude M has itself a magnitude M.

Figure S5a shows the behavior of $P(m|M)$, which progressively delocalizes towards large magnitude $m$ as $M$ increases. For comparison, the pdf corresponding to the ETAS model (Model 1), which does not depend on $M$, is plotted using a black dashed line.

Integrating $P(m|M)$ with $m$, we can estimate $S(m|M)$, the share of events with magnitude larger than $m$ that trigger an event of magnitude $M$ (see Figure S5b). If we consider, for instance, the median value $S(m|M)=0.50$, we notice that the corresponding triggering magnitude increases from $m=3.4$ for $M=3$, to $m=5.24$ for $M=7$. Thus, forecasting correctly half of the triggered events of magnitude $M$ requires detecting properly only the largest events if $M$ is large. However, it must be kept in mind that forecasting the background events of similar magnitude $M$ requires additional ingredients or mechanisms beyond model 4, which assumes them as Poissonian in time.

Using Eq.(8), we obtain a further diagnostic
$$\frac{P(m=M|M)}{P(m=M_0|M)} = \frac{\beta_1}{e^{(\beta_{bkg}-a-\beta_2)(M-M_0)}\left(\beta_1+\beta_2(e^{\beta_1(M-M_0)}-1)\right)} := R(M) \quad (11)$$

which quantifies the relative contributions of events of magnitude $M$ to those of magnitude close to the lower threshold $M_0$ in their propensity to trigger an event of magnitude $M$. Figure 1b plots the dependence of this ratio $R(M)$ as a function of $M$. One can observe that $R(M)<1$ for $M<M^*$ and $R(M)>1$ for $M<M^*$, where $M^*\approx 6.3$ for the parameters calibrated on the Californian catalog with Model 4. Thus, the triggering of earthquakes with $M<6.3$ is dominated by previous events with magnitude $M_0$, while the triggering of earthquakes with $M>6.3$ is dominated by previous events with magnitude $M$. This provides another confirmation of the major difference of Model 4 compared with the conclusions drawn using the standard ETAS model, where the triggering of events of any magnitude range is dominated by events with magnitude $M_0$, since $a - \beta_{bkg} < 0$ holds true in general.

In our calibration of the models, we considered that $M_0=M_c$, i.e., the minimum magnitude $M_c$ above which all events are thought to have been detected is equal to the minimum magnitude $M_0$ at which earthquakes can trigger other earthquakes. However, the true $M_0$ is likely much smaller than $M_c$ [5]. Since events with $M<M^*$ are either unpredictable background events or are mainly triggered by unobserved events, it follows that the performance of any statistical forecasting technique can be expected to be low for such a magnitude range. On the other hand, provided that $M^*> M_c$, events of magnitude $M>M^*$ are either background events or triggered by observable events of similar magnitude. Our best Model 4 should thus perform better on that magnitude range. One should, however, keep in mind that the qualitative behavior of $P(m|M)$ in other areas may be different, depending on the local set of exponents ($a$, $\beta_1$, $\beta_2$, $\beta_{bkg}$). For California, the cross-over magnitude $M^*\approx 6.3$ could be related to the rupture length equal to the dimension of its seismogenic crust [21].

**Earthquake doublets.** The mechanical and symmetry argument underlying Model 4 mentioned above supports the existence of earthquake doublets, i.e., observed pairs of earthquakes that are close in space and time and of similar magnitudes [22-24]. Indeed, Model 4 produces a reasonable fraction of doublets compared with the standard ETAS model or Model 1, as shown by numerical simulations performed as follows. Using the distribution of background events inverted in Models 1 and 4, we simulate two different catalogs using each of these two models, respectively. We keep only events with magnitude $\geq 6$ and select those that do not fall within the space-time window of a larger preceding event. This space-time window is set as +/-2 days and 2.5 times the source length of the first event. A pair of events A and B is then considered as a doublet if their magnitude difference is smaller than 0.5, they occur within 2 days, and the second event occurs within 2.5 times the source length of the first event. Using a catalog simulated using Model 1, we find that about 2% of events with $M \geq 6$ belong to a doublet, while about 15% of events with $M \geq 6$ belong to a doublet in a catalog simulated using Model 4. For California, the ANSS catalog features 30 events with $M \geq 6$, of which 27 remain after filtering according to the above criteria, and 4 of them (15%) belong to a doublet. As a result of the kinked GR law, Model 4 accounts for doublets, which appear to be a genuine feature of seismicity and not because of selection biases. Model 4 thus offers a quantitatively accurate estimation of magnitude correlations for the entire earthquake catalog while being formulated at the micro-scale of single triggering events. Eqs. (8) and (10) also allow us to estimate the 'bare' rate of doublets for Model 4, i.e., the number of doublets produced by direct triggering (while the estimates above are 'dressed' or 'renormalized' rates, obtained by considering the whole catalog, i.e., all generations along the triggering cascades). Getting rid of space and time windows, we simply integrate $P(m|M)$ with $m$ within *[M-0.5;M+0.5]*, and divide the result by the same quantity obtained using Model 1. Figure 1c shows that this doublet rate ratio is close to 1 when M is small and increases slightly faster than an exponential when $M \geq 4$. For instance, we see that the doublet rate ratio is about 12 for M=6. Estimations on the ANSS (dressed) catalog above provide a roughly similar ratio ~7.5 (a lower value that may be due to the use of finite space-time windows for the selection of doublets, while our triggering kernels extend infinitely in space).

**Conclusion.** We have compared four different stochastic models of seismicity and found that, for California, the model with parameters dependent on the magnitude of the triggering shock for the Omori time kernel and for the magnitude distribution of triggered events (model 4) offers the best performance in forecasting future seismicity. The agreement between the rate of doublets in synthetic and natural catalogs provides additional validation of this best model and, at the same time, supports the notion that earthquake doublets are a genuine feature of seismicity (and not spurious observations from selection bias). The ingredients of the best model can be rationalized from the norm and symmetry of the fluctuations of the stress and deformation tensors.

Model 4 predicts that earthquakes up to a critical magnitude $M^*$ (=6.3 for California) are preferably triggered by the smallest possible triggering shocks; in contrast, earthquakes with magnitude $M$ above $M^*$ are preferably triggered by events with a similar magnitude $M$. Thus, forecasting small events is inherently unfeasible due to a lack of observations of their (small magnitude) triggers. In contrast, a significant fraction of large events is triggered by large events. Being large, these triggering events are observed, and their catalog is close to complete for the global Earth seismicity. Thus, forecasting large events preceded by a set of seismic precursors [e.g., 25] would not be spoiled by the noise created by the myriad of small events whose recording is incomplete or whose parameters are highly uncertain. Given that only large events are relevant for risk assessment and protection, our results suggest that modeling the spatial anisotropy of sources of large spatial dimensions defines the next important conceptual step towards a successful operational earthquake forecasting system (as large earthquakes cannot be considered as point sources).

Our improved ability to forecast applies only to the subset of triggered earthquakes, while background events remain statistically unpredictable according to the investigated models. To improve, a better description of the background (non-triggered) part of the seismic catalogs is needed. This has been partially achieved in the present set of models by introducing a spatially heterogeneous structure for the set of background events. But there is no time dependence. A substantial increase of forecasting power should focus on modeling possible time variations of local or regional strain fields and other geophysical fields that could be correlated with seismicity [26].

**Supplementary Information for**

Triggering of large earthquakes is driven by their twins

Shyam Nandan[1], Guy Ouillon[2], Didier Sornette[3,4]

[1]Swiss Seismological Service, ETH Zürich, Sonneggstrasse 5, 8092, Zürich, Switzerland

[2]Lithophyse, 4 rue de l'Ancien Sénat, 06300 Nice, France

[3]Department of Management, Technology and Economics, ETH Zürich, Scheuchzerstrasse 7, 8092, Zürich, Switzerland

[4]Institute of Risk Analysis, Prediction and Management (Risks-X), Academy for Advanced Interdisciplinary Studies, Southern University of Science and Technology (SUSTech), Shenzhen, China

*Shyam Nandan

**Email:** shyam4iiser@gmail.com

**This PDF file includes:**

    Supplementary Text S1 – S8
    Figures S1 to S5
    Table S1
    SI References

## Supplementary Text

*All equations labelled with numbers refer to the main text; other equations are labelled with capital letters.*

### Text S1: Description of the models.

**Model 1** uses the conditional seismicity rate given by Eq. (1) but extends the standard ETAS model with a space-varying background intensity function given by:

$$\mu(x,y) = T^{-1} \sum_{i=1}^{N} IP_i \, \pi^{-1} \, Q \, D^{2Q}((x-x_i)^2 + (y-y_i)^2 + D^2)^{-1-Q} \tag{S1-1}$$

where $IP_i$ is the independence probability of the $i^{th}$ earthquake. It is not known a priori, but estimated iteratively using the Expectation Maximization (EM) algorithm used to invert the parameters (see Appendix SI Text S2). The proposed parameterization Eq. (S1-1) should not be confused with the triggering part described by Eq. (2) of the ETAS model, which also involves a summation over previous earthquakes. Here, the guiding idea is that the future background earthquakes occur mostly in regions where the intensity of past background earthquakes has been high.

**Model 2** also follows Eq. 1-2 and S1-3, but with a modified time kernel to $T_{norm}\{t - t_i + c(m_i)\}^{-p(m_i)} e^{-\frac{t-t_i}{\tau}}$, where $c(m_i) = c_0 10^{c_1 m_i}$ and $p = p_0 + p_1 m_i$ (Ouillon and Sornette, 2005; Nandan et al., 2021b; Sornette and Ouillon, 2005), i.e, the regularizer and the exponent of the time kernel feature exponential and linear dependence on the magnitude of the mainshock, respectively.

**Model 3** is defined by the following Eq. for the conditional seismicity rate of events of magnitude $m$:

$$\lambda(t,x,y,m|\mathcal{H}_t) = \mu(x,y) f_{bkg}(m) + \sum_{i:t_i<t} g(t-t_i, x-x_i, y-y_i, m_i) f_{aft}(m|m_i) \tag{S1-2}$$

where the background intensity function $\mu(x,y)$ is given by Eq. (S1-1) and the triggering function $g(t-t_i, x-x_i, y-y_i, m_i)$ is defined by Eq. (2). Thus, Model 3 differs from Model 1 by replacing Eq. (1) by Eq. (S1-2), which involves a new form of the GR law, which is not more factored out so as to express the existence of magnitude correlations. This new form is made of several parts: (i) $f_{bkg}(m)$ quantifies the magnitude distribution of the background earthquakes, given by

$$f_{bkg}(m) = \beta_{bkg} e^{-\beta_{bkg}(m-M_0)} \tag{S1-3}$$

(ii) $f_{aft}(m|m_i)$ generalizes the standard GR law and quantifies the conditional distribution of events of magnitude $m$ directly triggered by an earthquake of magnitude $m_i$, according to (Nandan et al, 2019a, Vere-Jones, 2005; Saichev and Sornette, 2005):

$$f_{aft}(m|m_i) = \begin{cases} C_1 e^{-(\beta_1)(m-M_0)} & \forall \, m < m_i \\ C_2 e^{-(\beta_2)(m-M_0)} & \forall \, m \geq m_i \end{cases} \tag{S1-4}$$

The constants $C_1$ and $C_2$ are derived from the constraints that $f_{aft}(m|m_i)$ is a probability density function (PDF), so that $\int_{M_0}^{\infty} f_{aft}(m|m_i) dm = 1$, and $C_2 e^{-(\beta_2)(m_i-M_0)} = C_2 e^{-(\beta_1)(m_i-M_0)}$, a condition that ensures the continuity of the PDF for $m = m_i$. Eq. (S1-4) implies that $f_{aft}(m|m_i)$ describes triggered earthquake magnitudes as clustered around that of their parent earthquake. Using nested likelihood tests, Nandan et al. (2019a) established the superiority of this model compared to the standard ETAS model with the standard GR law $e^{-\beta(m-M_0)}$, and speculated that these correlations between magnitudes of the triggering and triggered events can significantly improve the forecasting performance of the ETAS models.

***Model 4*** combines all previous novel ingredients of Models 2 and 3, i.e., it uses different magnitude distributions for triggered events and their trigger as in Model 3, and a time kernel dependent on the magnitude of the trigger as in Model 2.

**Text S2: Calibration of the models.**
In the following, we briefly describe the modifications brought to the Expectation Maximization process of inversion of Models 1-4 of the main text.
To invert the parameters of Model 1 and Model 2, we use the following algorithm (see Nandan et al., 2021a):
1. We start with an initial guess of the independence probability, $IP_i$, for all the earthquakes in the primary catalog. This initial guess can be created by simply drawing a uniform random number between 0 and 1 corresponding to each earthquake. We also make an initial guess for the parameters of the triggering kernel as well as the smoothing parameters ($D, Q$) for the background kernel (Eq. S1-1).
2. E step: Using the current value of the parameters and the independence probabilities, we define the probability that the $i^{th}$ earthquake triggered the $j^{th}$ earthquake as:

$$P_{ij} = \frac{g(t_j - t_i, x_j - x_i, y_j - y_i, m_i)}{\mu(x_j, y_j) + \sum_{i:t_i<t_j} g(t_j - t_i, x_j - x_i, y_j - y_i, m_i)} \quad (S2-1)$$

where, $\mu(x_j, y_j) = T^{-1} \sum_{i \neq j} IP_i \, \pi^{-1} \, Q \, D^{2Q} \left((x_j - x_i)^2 + (y_j - y_i)^2 + D^2\right)^{1+Q}$.

The reason for the choice of the summation $\sum_{i \neq j}$ will become apparent in the following steps. The new estimates of the independence probabilities can be obtained as $IP_j = 1 - \sum_i P_{ij}$. We update the current estimates of independence probabilities with the new estimates.

3. M step 1: Using the current estimates of all $IP_i$ values, we define the PDF of the location of background earthquakes at the location of $j^{th}$ background earthquake as:

$$\mu_{PDF}(x_j, y_j) = \frac{\sum_{i \neq j} IP_i \, \pi^{-1} \, Q \, D^{2Q} \left((x_j - x_i)^2 + (y_j - y_i)^2 + D^2\right)^{1+Q}}{\sum_{i \neq j} IP_i} \quad (S2-2)$$

Using $\mu_{PDF}(x_j, y_j)$ and $IP_j$, we can define the complete data log-likelihood for the spatial distribution of the background earthquakes as:

$$LL_{bkg} = \sum_{j=1}^{N} IP_j \times \ln \mu_{PDF}(x_j, y_j) \quad (S2-3)$$

$LL_{bkg}$ can be optimized for the parameters $D$ and $Q$ to obtain their new estimates. Several caveats are important to consider:
   a. The minimum value of $D$ is set to the location error.
   b. To obtain $\mu_{PDF}(x_j, y_j)$, the summation in the right-hand side of the Equation is done for $i \neq j$. Otherwise, the optimization of $LL_{bkg}$ leads to D being very close to 0 and $Q$ assuming very large values, which creates a Dirac function at the location of the $j^{th}$ earthquake, thus leading to the maximal possible value of $LL_{bkg}$. This artificial situation can be avoided if one uses all earthquakes except the $j^{th}$ earthquake to explain the background rate at its location, which amounts to a leave-one-out strategy.

4. M step 2: Maximize $LL_{trig}$, defined in Equation S2-4, for the parameters ($\theta$) of the triggering kernel.
$LL_{trig} = \sum_i \{-\log(\Gamma(\psi_i + 1)) - G_i(\theta) + \psi_i \log(G_i(\theta))\} + \sum_{ij} P_{ij}^k \log \{g(t_j - t_i, x_j - x_i, y_j - y_i, m_i)\}$ (S2-4)

Where, $\psi_i = \sum_{j \, \forall t_i < t_j} P_{ij}$, is the number of direct aftershocks of the $i^{th}$ earthquake and $G_i(\theta)$ is the expected number of offsprings triggered by an earthquake $(t_i, x_i, y_i, m_i)$ within the study region $S$ and the primary time period $[T_1, T_2]$ and is given by $\int_{\max(t_i,T_1)}^{T_2} \iint_S g(t - t_i, x - x_i, y - y_i, m_i)\, dx\, dy\, dt$.

5. Update the current estimates of all the parameters to the new estimates obtained in steps 3 and 4.
6. Repeat steps 2-5 until convergence.

For Models 3 and 4, we follow the same algorithm as above with the modification that the triggering probabilities are now defined as:

$$P_{ij} = \frac{g(t_j - t_i, x_j - x_i, y_j - y_i, m_i) f_{aft}(m_j|m_i)}{\mu(x_j, y_j) f_{bkg}(m_j) + \sum_{i: t_i < t_j} g(t_j - t_i, x_j - x_i, y_j - y_i, m_i) f_{aft}(m_j|m_i)} \quad (S2-5)$$

For these two models the parameters of $f_{bkg}$ and $f_{aft}$ defined in Equations 5-6 are jointly maximized on the M step 2 of the algorithm prescribed above. For optimization of parameters of $f_{bkg}$ and $f_{aft}$ and additional log-likelihood term, $LL_+$, is added to $LL_{trig}$ defined in Equation (S2-4).

$$LL_+ = \log(\beta_{bkg}) \sum_j IP_j - \beta_{bkg} \sum_j IP_j(m_j - M_0) + \sum_{ij} P_{ij} \log f_a(m_j|m_i, \beta_1, \beta_2) \quad (S2-6)$$

where,

$$\log f_a(m|m_i, \beta, \delta) = \begin{cases} -\log\left[\frac{2\delta}{\beta^2 - \delta^2}\left\{\left(\frac{\beta + \delta}{2\delta}\right) e^{-(\beta-\delta)M_0} - e^{-(\beta-\delta)m_i}\right\}\right] - (\beta - \delta)m & \forall m \leq m_i \\ -\log\left[\frac{2\delta}{\beta^2 - \delta^2}\left\{\left(\frac{\beta + \delta}{2\delta}\right) e^{-(\beta-\delta)M_0} - e^{-(\beta-\delta)m_i}\right\}\right] + 2\delta m_i - (\beta + \delta)m & \forall m > m_i \end{cases}$$

; $\beta = \frac{\beta_1 + \beta_2}{2}$ and $2\delta = \beta_2 - \beta_1$.

### Text S3: Description of the data set.

We use ~1.2 million reported earthquakes (with $M \geq 1$) within 1975-2020 in study region (co-ordinates of the spatial polygon in Table S1) surrounding California in the ANSS earthquake catalog. Nandan et al. (2021a) have shown using several proxies that the catalog within this region and time period can be considered to be reasonably complete above M = 3. Thus, we set magnitude of completeness, $M_c = 3$ in this study as well. All earthquakes above $M_c$ will be used to invert the parameters of the models, as well to simulate future datasets. We shall also assume that $M_c$ coincides with $M_0$. However, only events above $M_t$ (chosen by the user) will be used to check the forecasts.

An important consideration when calibrating the ETAS model is the choice of the auxiliary and primary periods. The earthquakes in the auxiliary period only serve as sources. In contrast, the primary period's earthquakes can act as both sources and targets during the ETAS model's calibration. Without this consideration, the calibration process would yield a disproportionate fraction of background earthquakes at the beginning of the catalog, as there are no or very few events to act as potential triggers. For the catalog used in this study, the earthquakes between 1975 and 1981 are taken as part of the auxiliary catalog. All earthquakes following 1981 are taken as part of the primary catalog.

### Text S4: Identification of the best model.

Following the norm of scientific epistemology, we propose to identify the best model for the understanding of seismic processes as the one that provides the best predictive ability. For this, we set up pseudo-prospective forecasting experiments using earthquakes reported in the Advanced National Seismic System (ANSS) catalog around the state of California, for which we estimated the completeness magnitude $M_c$=3 (see SI Appendix, Text S3 for the description of the dataset used in this study). In these experiments, we use the early part of the data to calibrate the models and leave the future data as unseen to compare with the forecasts constructed using the first part of the data. Starting on January 1, 1990, we perform 368 pseudo prospective experiments. The duration of testing periods is fixed to 30 days, and all the testing periods are non-overlapping.

The four competing models issue forecasts as simulated stochastic catalogs (location, occurrence time and magnitude of future events). Each model simulates 500,000 catalogs for each of the testing periods to obtain high-precision distributions of forecasts, following the recommendation of Nandan et al (2019b, 2019c). In total, we perform 184,000,000 simulations for creating the forecasts of a given model for all the 368 testing periods, and this for each of the four models. The stochastic catalogs are used to construct the models' forecasts at any spatial resolution and magnitude threshold during the testing periods. In this work, the models are evaluated at five different spatial resolutions: $97\ km^2$, $4 \times 97\ km^2$, $4^2 \times 97\ km^2$, $4^3 \times 97\ km^2$, $4^4 \times 97\ km^2$, and five different target magnitude thresholds ($M_t$): 3, 3.5, 4, 4.5, 5. During a given testing period, competing models forecast the distribution of the number of earthquakes (of magnitude $\geq M_t$) in triangular pixels of equal area, which are used to divide the study region. Events of lower magnitude are used in the calibration and simulations, but are not part of the test sets. We then count the actual number of observed earthquakes $\geq M_t$ within each pixel. With these two pieces of information, the log-likelihood $LL_A^i$ of Model A during the $i^{th}$ testing period is defined by $LL_A^i = \sum_{j=1}^{N} \ln[Pr_{A,j}^i(n_j^i)]$, where $Pr_{A,j}^i$ is the PDF of the number of earthquakes $\geq M_t$ forecasted by Model A in pixel $j$ during the $i^{th}$ testing period, while $n_j^i$ is the observed number of such earthquakes in the same pixel and time period. Using the likelihoods $LL_A^i$ and $LL_B^i$ of two competing models A and B, we can then define the information gain ($IG_{AB}^i$) of Model A over Model B during the $i^{th}$ testing period as $IG_{AB}^i = LL_A^i - LL_B^i$. Such information gains are computed at each spatial resolution.

As a repeated calibration of the four models on progressively increasing size of the training data, a time series is obtained for the inverted parameters for all the four models, presented in SI Appendix Text S5.

Before ranking the models by their performance in the pseudo-prospective tests, we first establish in Appendix SI Text S6 that Model 1 is a reliable benchmark. We then compare the performance of Models 2-4 relative to Model 1, treated as the null model.

Figure S4 shows the mean information gain ($MIG = \frac{\sum_i IG_i}{368}$) that models 2-4 (shown as blue, orange, and red colors, respectively) obtain with respect to Model 1 for magnitude threshold ($M_t$=3) of the testing catalog at different spatial resolutions (while all events above $M_c$ are used for model inversion). Figure S3 shows the MIG of models 2-4 with respect to Model 1 at all the magnitude thresholds. In total, we did the evaluations of the models with respect to Model 1 at 25 settings (resulting from a combination of five different $M_t$ thresholds and five different spatial resolutions). One can observe from Figure S5 (and Figure S3) that models 2 and 4 significantly outperform Model 1 in nearly all testing settings, suggesting that a magnitude-dependent Omori kernel is a first order feature of a correct modelling of seismicity (see SI Appendix Text S7 for a discussion of the rather peculiar performance of Model 3).

### Text S5: Time-dependence of the estimated parameters for the four models.

Figure S1 presents the time series obtained from the calibration of the four models on the training sets associated with each of the 368 successive testing periods. The size of these training sets is thus progressively increasing as more and more earthquakes are added to the training earthquake catalog.

We notice that both the regularizer and the exponent of the time kernel show strong dependence on the magnitude $m$ of the trigger in Models 2 and 4. For instance, the regularizer increases as $10^{-3.94+0.34m}$ and $10^{-3.91+0.33m}$ for Models 2 and 4, respectively. Note that it increases more slowly than the rupture length, suggesting that this characteristic time does not scale with rupture duration. For the final training period, the exponent of the Omori law increases as $0.46 + 0.15m$ for both models, i.e. in a manner very similar to previous results reported in (Sornette and Ouillon, 2005; Ouillon and Sornette, 2005; Ouillon et al., 2009; Tsai et al., 2012; Nandan et al., 2021b) for many regional and global catalogs.

We also notice that Models 3 and 4 infer a substantial kink in the magnitude distribution of triggered events, with the kink appearing at the magnitude of the trigger (Eq. S1-4). Below that magnitude, the magnitudes of the triggered earthquakes follow an exponential distribution with an exponent (1.6 for both models for the final training period) that is much smaller than that of the background earthquakes (2.52 and 2.40 Models 3 and 4, respectively, over the same period). The magnitudes of triggered events larger than their trigger also follow an exponential distribution but with an exponent (3.12 for both models for the final training period) comparatively much larger than that of the background earthquakes. For Models 1 and 2, we show the time series of $\beta$ (defined in the standard GR law $\beta e^{-\beta(m-M_0)}$ in Eq. 1), which characterizes the magnitude distribution of both the background and triggered earthquakes. As all models use the same training catalog, these estimates of $\beta$ are also valid for Models 3 and 4. For the final training period, $\beta$=

2.39, quite close to the average estimate of $\beta_{bkg}$ for Model 4 and only marginally smaller than the estimate of $\beta_{bkg}$ for Model 3.

All parameters show variations with time to some extent. A major source of this time variation is a progressive increase in the size of the training catalog, leading to a converging trend in some of the parameters such as $\mu$, the $\beta$s, branching ratio and $p_1$. However, in the time series of some of the parameters such as $K$, $a$, $c_0$, $c_1$, $p_0$ and so on, one can also notice a tendency for sudden, yet small, jumps. These jumps are associated with some prominent events in the earthquake catalog, including the Landers, Hector Mine, El-Major earthquakes and so on.

## Text S6 Performance of Model 1 compared to a spatially and temporally homogenous Poisson process (STHPP).

We first compare Model 1 relatively to a spatially and temporally homogenous Poisson process (STHPP) to establish it as a reliable benchmark against which all the other models will be evaluated. As the name suggests, the STHPP model forecasts the rate of future earthquakes as being homogenous in space and uniform in time. The rate forecasted by the STHPP is estimated using the data in the training period and is given by:

$$\lambda = \frac{N(\geq M_t)}{A \times T} \qquad (S6-1)$$

where $\lambda$ is the average rate of earthquakes with magnitude larger than $M_t$ per day per $km^2$, $A$ is the area of the study region, and $T$ is the training catalog's time duration. $M_t$ is equal to the testing magnitude threshold. For the current study region, $A = 961,238\ km^2$. The value of $T$ depends on the starting time of the testing period (which is also the end time of the training dataset). For the first testing period, $T = 3,286\ days$ (~9 years, considering training data between January 1, 1981, and January 1, 1990), increasing by a constant step of 30 days as the training period becomes larger and larger. Having obtained $\lambda$ from a given training period, the forecast of the STHPP model for the following testing period (of duration 30 days) is prescribed as a mean rate in all the equal-area pixels as $\lambda_{pixel} = \lambda \times 30 \times A_{pixel}$, where $A_{pixel}$ is the area of triangular pixels with which the study region is tiled. For the $i^{th}$ testing period, the performance of the model is estimated using the Poissonian log-likelihood as:

$$LL_{STHPP}^i = \sum_j n_j^i \ln \lambda_{pixel}^i - \lambda_{pixel}^i - \ln n_j^i! \qquad (S6-2)$$

Index *i* stands for the testing period, while *j* stands for the pixel's index. The information gain of Model 1 over STHPP in the $i^{th}$ testing period is simply $IG_i = LL_{Model\ 1}^i - LL_{STHPP}^i$. We then obtain the information gain per earthquake (IGPE) that Model 1 obtains over STHPP as $IGPE = \frac{\sum_i IG_i}{\sum_i N_i(\geq M_t)}$, where $N_i(\geq M_t)$ is the observed number of earthquakes with magnitude larger than $M_t$ during the $i^{th}$ testing period. Figure S2 shows the IGPE (colored bars) and its 95% confidence interval (orange error bars obtained using bootstrapping the IG's obtained from the 368 testing periods). Since the models are compared at five different spatial resolutions (indicated on the x-axis) and five testing magnitude thresholds (indicated by colors of the bars), we get five groups of five bars. Figure S2 leads to the following conclusions:

1. At all spatial resolutions, Model 1 significantly outperforms the STHPP with IGPE ranging from 1.72 ($A_{pixel}$ = 24,905 $km^2$, $M_t = 5$) to 4.3 ($A_{pixel}$ = 97 $km^2$, $M_t = 3$) with $p-values$ resulting from pairwise T-tests being below the computer precision.

2. With the increasing spatial resolution (i.e., decreasing $A_{pixel}$), Model 1 obtains higher IGPE over the STHPP.

3. At all the spatial resolutions, although there is no general trend of a drop in performance with increasing $M_t$, Model 1 features a consistently lower IGPE at $M_t = 5$ compared to its IGPE for other $M_t$'s. However, we cannot present evidence whether IGPE could continue decreasing at larger $M_t$ values, as the number of simulated events would be too small to reliably estimate the performance of the forecasts. For that purpose, one would have to drastically increase the number of simulated catalogs.

The IGPEs of Model 1 over the STHPP translate into a probability gain ($= e^{IGPE}$) ranging between 5.62 and 72.96. These numbers essentially indicate how likely Model 1 is to explain an individual earthquake relative to the STHPP. These numbers point that Model 1 is a much superior model to STHPP and can thus act as a strong benchmark against which all other models can be evaluated.

**Text S7 Discussion of the performance of Model 3.**
The performance of Model 3 with respect to Model 1 behaves peculiarly. It underperforms at the lower target magnitude thresholds and then starts to outperform Model 1 at higher magnitude thresholds. One possible explanation of this systematic behavior is the following: Model 3 is parametrized with a distribution of the triggered events dependent on the magnitude of their trigger. These distributions show a kink in the Gutenberg Richter distribution strictly at the magnitude of the trigger, with the implication that an earthquake of a certain magnitude has a higher penchant for triggering aftershocks of similar magnitude compared to the pure Gutenberg-Richter case (as assumed in Models 1 and 2). In a way, an earthquake of a given magnitude contributes more to the triggering of similar magnitude events, if not dominating their triggering (see below). We also know from the parametrization of model 3 that the Omori exponent, which controls the triggered events decay rate, is the same for all events. However, the outperformance of Model 2 over Model 1 indicates that the magnitude dependence of the Omori exponent is an important feature that we should not ignore. However, in Model 3, the Omori exponent is appropriate only for a single trigger magnitude, i.e., the magnitude at which the underlying real *p(m)* law intersects with the inverted constant *p*, which happens for *M~4*. For lower magnitudes triggers (M=3-4), which contribute to the triggering of similar magnitude targets, the model would predict a too strong clustering in time than due to the underlying *p(m)* model, thus leading to a worse forecast. As the targets' magnitude threshold increases (M=4-5), lower triggers magnitudes' contribution diminishes, and triggers with magnitudes (M=4-5) start contributing significantly. For these events, the constant Omori exponent is more in agreement with the values that would be provided by *p(m)*, which serendipitously leads to an improved forecast. This explanation then forecasts that at a higher target magnitude threshold, Model 3 performance should further diminish.

**Text S8 Contribution of mainshocks of different magnitudes to the triggering of aftershocks above some magnitude.**
Here we derive expressions (4) and (5) that form the basis of equation (6) and of Figure 1 in the main text, Let $m_{main}$ be the magnitude of a triggering event. The magnitude distribution of its directly triggered events is given by

$$f_{aft}(m) = \begin{cases} C_1 e^{-(\beta_1)(m-M_0)} & \forall\, m < m_{main} \\ C_2 e^{-(\beta_2)(m-M_0)} & \forall\, m \geq m_{main} \end{cases} \quad (S8\text{-}1)$$

where $C_1$ and $C_2$ are derived from the constraints that $f_{aft}(m)$ is a probability density function (PDF), so that $\int_{M_0}^{\infty} f_{aft}(m)dm = 1$, and $C_1 e^{-(\beta_1)(m_{main}-M_0)} = C_2 e^{-(\beta_2)(m_{main}-M_0)}$, a condition ensuring the continuity of the PDF for $m = m_{main}$. Using these constrains, we find:

$$C_1 = C_2 e^{(\beta_1-\beta_2)(m_{main}-M_0)}$$

$$C_2 = \left( \frac{e^{(\beta_1-\beta_2)(m_{main}-M_0)}\left(1 - e^{-\beta_1(m_{main}-M_0)}\right)}{\beta_1} + \frac{e^{-\beta_2(m_{main}-M_0)}}{\beta_2} \right)^{-1} \quad (S8-2)$$

In addition, we know from the productivity law that the total number of aftershocks above $M_0$ triggered by an event of magnitude $m_{main}$ is given by $Ke^{a(m_{main}-M_0)}$.

The cumulative number of aftershocks triggered above a threshold $m_{aft}$ is the fraction $I_{m_{main}}(m_{aft})$ of $Ke^{a(m_{main}-M_0)}$ aftershocks that have their magnitudes larger than $m_{aft}$, where

$$I_{m_{main}}(m_{aft}) = \int_{m_{aft}}^{\infty} f_{aft}(m)dm \quad (S8-3)$$

For $m_{aft} < m_{main}$,

$$I_{m_{main}}(m_{aft}) = \int_{m_{aft}}^{\infty} f_{aft}(m)dm = \int_{m_{aft}}^{m_{main}} f_{aft}(m)dm + \int_{m_{main}}^{\infty} f_{aft}(m)dm$$

which gives

$$I_{m_{main}}(m_{aft}) = \frac{C_1}{\beta_1}(e^{-\beta_1(m_{aft}-M_0)} - e^{-\beta_1(m_{main}-M_0)}) + \frac{C_2}{\beta_2}e^{-\beta_2(m_{main}-M_0)} \quad \text{for } m_{aft} < m_{main},$$
(S8-4)

And from (S8-2), this simplifies into
$$I_{m_{main}}(m_{aft}) = 1 - \frac{C_1}{\beta_1}(1 - e^{-\beta_1(m_{aft}-M_0)}) \quad \text{for } m_{aft} < m_{main}, \quad \text{(S8-5)}$$

The number of aftershocks of magnitudes $m_{aft} < m_{main}$ triggered by a mainshock of magnitude $m_{main}$ is thus

$$Ke^{a(m_{main}-M_0)} I_{m_{main}}(m_{aft}) = K e^{a(m_{main}-M_0)} \left[1 - \frac{C_1}{\beta_1}(1 - e^{-\beta_1(m_{aft}-M_0)})\right]$$

$$= K e^{a(m_{main}-M_0)} \left[1 - \beta_2 \frac{1}{1-(1-\frac{\beta_1}{\beta_2})e^{-\beta_1(m_{main}-M_0)}} (1 - e^{-\beta_1(m_{aft}-M_0)})\right] \quad \text{(S8-6)}$$

by using expression (S8-2) for $C_1$. This shows that the productivity law is changed from the unconditional form $Ke^{a(m_{main}-M_0)}$ (i.e. for all possible aftershock magnitudes) to

$$1 - \beta_2 \frac{1}{1-(1-\frac{\beta_1}{\beta_2})e^{-\beta_1(m_{main}-M_0)}} (1 - e^{-\beta_1(m_{aft}-M_0)}) \quad \text{for } m_{aft} < m_{main} \quad \text{(S8-7)}$$

for aftershocks of magnitudes larger than $m_{aft}$, in the case where $m_{aft} < m_{main}$.

For $m_{aft} \geq m_{main}$, $\quad I_{m_{main}}(m_{aft}) = \frac{C_2}{\beta_2}e^{-\beta_2(m_{aft}-M_0)}.$ (S8-8)

Consider the case in which $m_{aft} = M_0$. In this case, $m_{main}$ will always be greater than $m_{aft}$ by definition. The total number of aftershocks of magnitude larger $M_0$ than scales with mainshock magnitude as $Ke^{a(m_{main}-M_0)} * (1 - \frac{C_1}{\beta_1}(1 - e^{-\beta_1(m_{aft}-M_0)})) = Ke^{a(m_{main}-M_0)}$, which retrieves the total average number of events triggered by a mainshock of magnitude $m_{main}$.

For $m_{aft} \geq m_{main}$, replacing the expression of $C_2$ in expression (S8-2) yields

$$I_{m_{main}}(m_{aft}) = \beta_1 \frac{e^{(\beta_2-\beta_1)(m_{main}-M_0)}}{1-(1-\frac{\beta_1}{\beta_2})e^{-\beta_1(m_{main}-M_0)}} e^{-\beta_2(m_{aft}-M_0)} \quad \text{(S8-9)}$$

The number of aftershocks of magnitudes $m_{aft} \geq m_{main}$ triggered by a mainshock of magnitude $m_{main}$ is thus

$$I_{m_{main}}(m_{aft}) Ke^{a(m_{main}-M_0)} = K e^{a(m_{main}-M_0)} \beta_1 \frac{e^{(\beta_2-\beta_1)(m_{main}-M_0)}}{1-(1-\frac{\beta_1}{\beta_2})e^{-\beta_1(m_{main}-M_0)}} e^{-\beta_2(m_{aft}-M_0)}$$

$$= K \beta_1 \frac{e^{(a+\beta_2-\beta_1)(m_{main}-M_0)}}{1-(1-\frac{\beta_1}{\beta_2})e^{-\beta_1(m_{main}-M_0)}} e^{-\beta_2(m_{aft}-M_0)} \quad \text{(S8-10)}$$

This shows that the productivity law is changed from the unconditional form $Ke^{a(m_{main}-M_0)}$ (i.e. for all possible aftershock magnitudes) to

$$K\beta_1 \frac{e^{(a+\beta_2-\beta_1)(m_{main}-M_0)}}{1-(1-\frac{\beta_1}{\beta_2})e^{-\beta_1(m_{main}-M_0)}} \quad \text{for } m_{aft} \geq m_{main} \quad \text{(S8-11)}$$

for aftershocks of magnitudes larger than $m_{aft}$, in the case where $m_{aft} \geq m_{main}$. Note that this expression (S8-10) reduces to K for $m_{main} = M_0$ as it should.

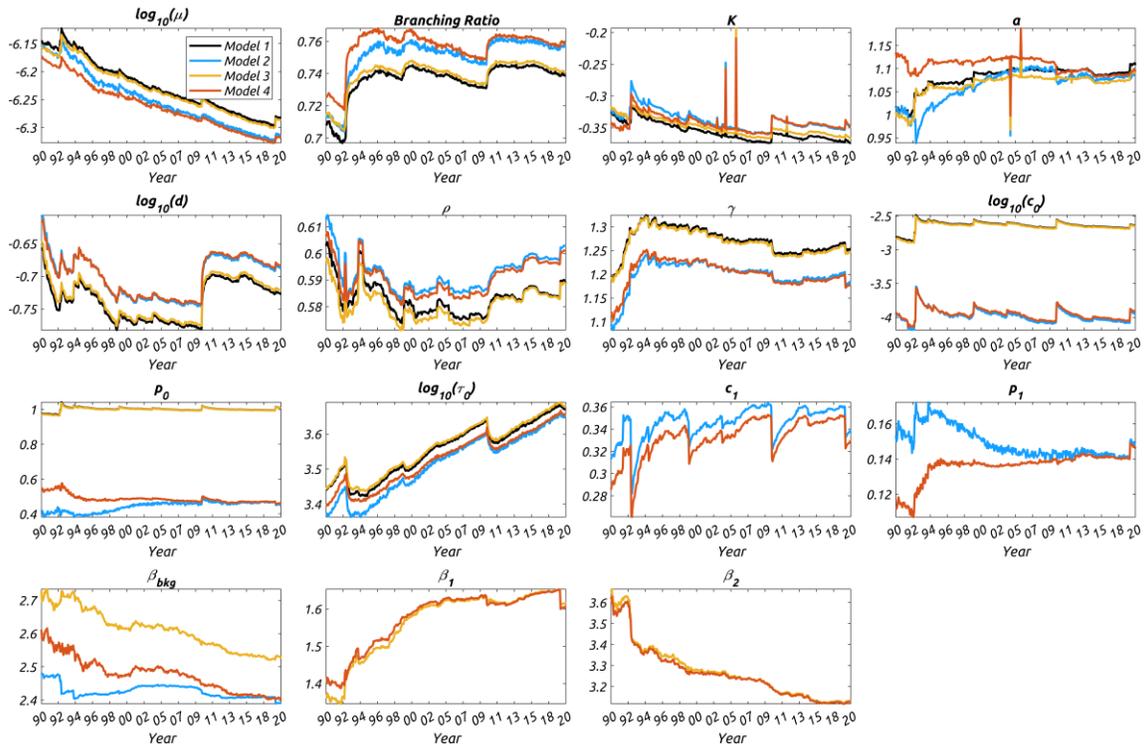

**Fig. S1.** Time series of parameters of the four models represented using four colors. Parameters $c_1$ and $p_1$ are only present in Models 2 and 4, and parameters $\beta_{bkg}$, $\beta_1$ and $\beta_2$ are only present in models 3 and 4.

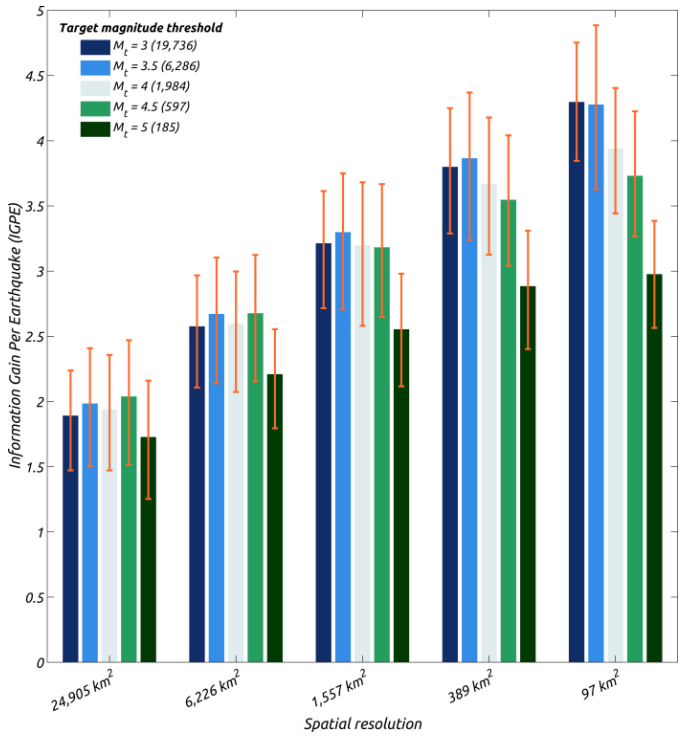

**Fig. S2.** Information gain per earthquake (IGPE) in colored bars and its 95% confidence interval (orange error bars) of Model 1 compared with the spatially and temporally homogenous Poisson process (STHPP). These bars and statistics are obtained using bootstrapping the IGPE's obtained from the 368 testing periods.

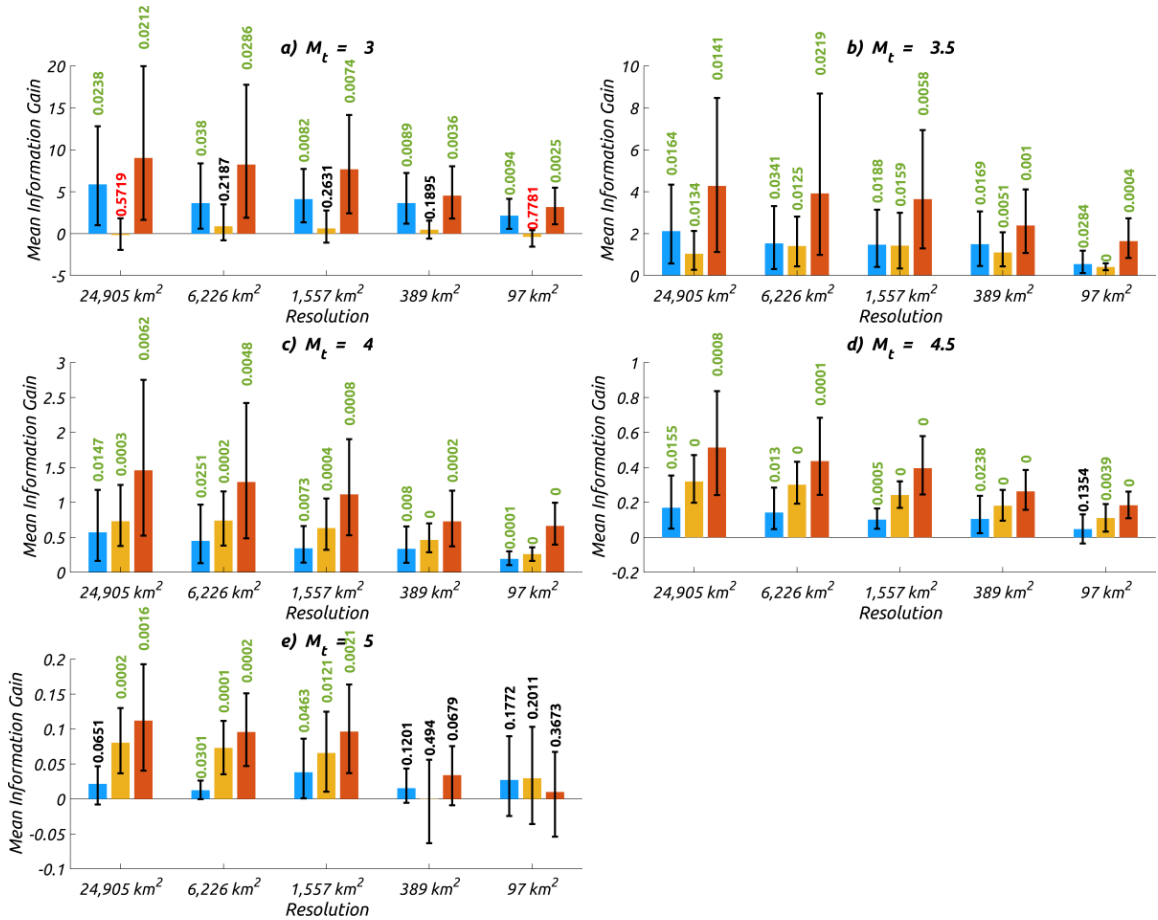

**Fig. S3.** Mean information gain (MIG) that Models 2-4 obtain with respect to Model 1 for 25 different testing settings; the five panels correspond to the five magnitude thresholds ($M_t$) of target catalog at which the testing was performed; the x-axis shows the five spatial resolutions at which models were compared; error bars show the 95% confidence interval of the MIG obtained using bootstrapping; the number above each bar is the p-value resulting from testing the null hypothesis that each of the MIGs are equal to 0 against the alternative that they are significantly larger than 0, using the paired T-test; Green, black or red color of the numbers indicate that the null hypothesis can be rejected (at a significance level of 0.05), not rejected or that the null model is significantly more informative than the alternative model, respectively.

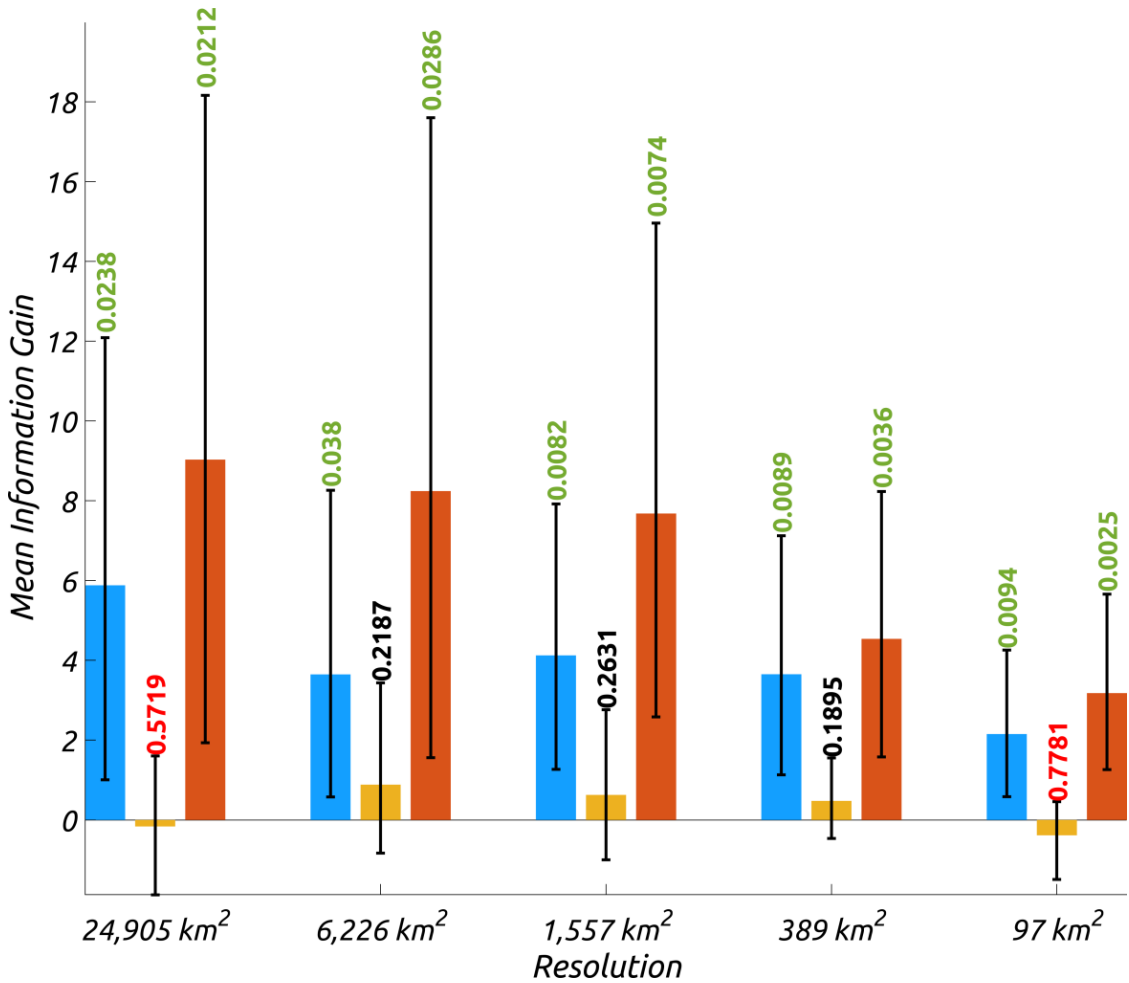

**Figure S4.** Mean information gain (MIG) that models 2-4 obtain with respect to Model 1 for magnitude threshold ($M_t$) of target catalog equal to 3; the x-axis shows the five spatial resolutions at which models were compared; error bars show the 95% confidence interval of the MIG obtained using bootstrapping; the number above each bar is the p-value resulting from testing the null hypothesis that each of the MIGs are equal to 0 against the alternative that they are significantly larger than 0, using the paired T-test; Green, black or red color of the numbers indicate that the null hypothesis can be rejected (at a significance level of 0.05), not rejected or that the null model is significantly more informative than the alternative model, respectively.

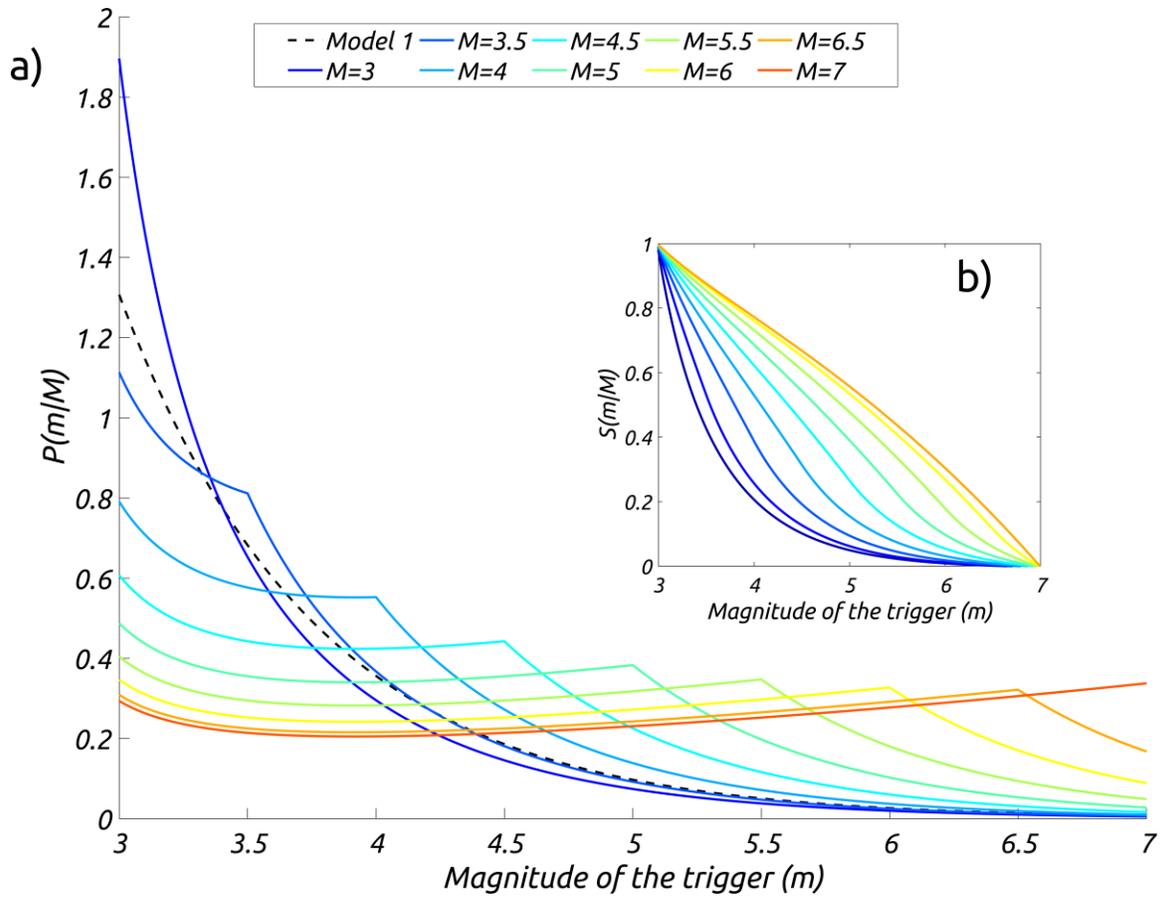

**Fig. S5. (a)** Conditional probability that an event of magnitude M has been triggered by an event of magnitude m, P(m|M). The black dashed line corresponds to Model 1. **(b)** Conditional probability that an event of magnitude M has been triggered by an event of magnitude m or larger, S(m|M). In both panels, different colors correspond to different values of magnitude of target (M), indicated in the legend; The curves are obtained for the parameters of Model 4 given by $a = 1.1, \beta_1 = 1.6, \beta_2 = 3.1, \beta = 2.4, M_0 = 3$, which best fit the ANSS Californian catalog.

**Table S1.** Spatial polygon corresponding to the study region.

| Longitude | Latitude |
|---:|---:|
| -125.7 | 43.5 |
| -118.5 | 43.5 |
| -118.5 | 39.7 |
| -113.6 | 36.1 |
| -112.6 | 34.6 |
| -112.6 | 34.3 |
| -113.1 | 32.7 |
| -113.2 | 31.8 |
| -114.5 | 31.2 |
| -117.1 | 31 |
| -117.4 | 31.1 |
| -118.3 | 31.5 |
| -118.8 | 32.4 |
| -121.3 | 33.3 |
| -122 | 34 |
| -124.3 | 37.5 |
| -125.9 | 40 |
| -125.9 | 40.5 |
| -125.7 | 43 |
| -125.7 | 43.5 |